\documentclass[a4paper]{jpconf}
\usepackage{graphicx}
\usepackage{amsmath}
\usepackage{amssymb}
\usepackage{citesort}
\begin{document}
\title{\boldmath$H^\pm W^\mp$ production in the MSSM at the LHC}
\author{David~Eriksson$^1$, Stefan~Hesselbach$^{2,}$\footnote[3]{Speaker} and Johan~Rathsman$^1$}
\address{$^1$ High Energy Physics, Uppsala University,
Box 535, S-75121 Uppsala, Sweden}
\address{$^2$ School of Physics \& Astronomy, University of Southampton,
Southampton SO17~1BJ, UK}
\ead{s.hesselbach@phys.soton.ac.uk}
\begin{abstract}
We investigate the viability of observing charged Higgs bosons ($H^\pm$)
produced in association with $W$ bosons at the CERN Large Hadron Collider, 
using the leptonic decay $H^+ \to \tau^+ \nu_\tau$ and hadronic $W$ decay,
within the Minimal Supersymmetric Standard Model.
Performing a parton level study we show how the irreducible Standard Model 
background from $W+2$ jets can be controlled by applying appropriate cuts. 
In the standard $m_h^\mathrm{max}$  scenario we find a viable signal 
for large $\tan\beta$ and intermediate $H^\pm$ masses ($\sim m_t$).
\end{abstract}

The quest for understanding electroweak symmetry breaking and mass generation 
is one of the driving forces behind the upcoming 
experiments at the CERN Large Hadron Collider (LHC). 
In the Minimal Supersymmetric Standard Model (MSSM),
which is a two Higgs Doublet Model (2HDM) of type II, the
Higgs sector consists of three neutral and one charged Higgs bosons
after electroweak symmetry breaking.
The charged Higgs boson ($H^\pm$) is of special interest since its discovery 
would constitute an indisputable proof of physics beyond the Standard Model.

The main production mode of charged Higgs bosons at hadron colliders is in
association with top quarks through the $gb \to H^-t$
and $gg \to H^-t\bar{b}$ processes~\cite{Barnett:1987jw,Bawa:1989pc,Borzumati:1999th,Miller:1999bm} with the
former one being dominant for heavy charged Higgs bosons 
$m_{H^\pm} \gtrsim m_t$ and the latter one for light ones 
$m_{H^\pm} \lesssim m_t - m_b$. Recently a new method for matching 
the differential cross-sections for the two production modes has been 
developed~\cite{Alwall:2004xw} resulting in a significantly improved 
discovery potential in the transition region $m_{H^\pm} \sim m_t$ 
\cite{Alwall:2007}.

A complementary production mode of charged Higgs bosons is in
association with $W$-bosons. Although the production 
cross-section~\cite{Dicus:1989vf,BarrientosBendezu:1998gd}
is large, an earlier study~\cite{Moretti:1998xq} 
using the hadronic charged Higgs decay,
$H^+ \to t \bar{b}$, came to the conclusion that the signal is overwhelmed by
the $t \bar t$ background. 
Here we report results of our study~\cite{Eriksson:2006yt} of the prospects of using instead
the $H^+ \to \tau^+ \nu_\tau$ decay together with $W \to 2 \textrm{ jets}$.

The dominant production mechanisms for $H^\pm W^\mp$ at hadron colliders 
are $b\bar{b}$ annihilation at tree-level and gluon fusion 
at one-loop-level.
In this study we focus on the parameter region with intermediate 
$H^\pm$ masses ($\sim m_t$)
and large $\tan\beta$, where 
the decay $H^\pm \rightarrow \tau\nu_\tau$ has a large branching ratio and 
where the $b\bar{b}$ annihilation dominates. 

%\begin{figure}[h]
%\begin{minipage}{10pc}
%\centering
%\includegraphics[width=10pc]{feynbbHW-s-fig.eps}
%\end{minipage}\hspace{1pc}%
%\begin{minipage}{8pc}
%\includegraphics[width=8pc]{feynbbHW-t-fig.eps}
%\end{minipage}\hspace{2pc}%
%\begin{minipage}[t]{17pc}
%\caption{Feynman diagrams for $H^\pm W^\mp$ production via
%$b\bar{b}$ annihilation.}\label{feynsignal}
%\end{minipage}
%\end{figure}

We have implemented~\cite{DEriksson} the two processes $b\bar b \rightarrow H^+ W^-$ 
and $b\bar b \rightarrow H^- W^+$ as separate external processes to
{\sc Pythia}~\cite{Sjostrand:2000wi}.
In principle, the implementation in {\sc Pythia} makes a generation of
the complete 
final state possible, but for this first study we have chosen to stay
on leading 
order parton level.
For the calculation of the MSSM scenario and the corresponding Higgs
masses and the branching ratios of $H^\pm$ 
we use {\sc FeynHiggs} 2.2.10 \cite{Hahn:2005cu}.
For simplicity we only consider hadronic decays of the $\tau$-lepton, 
$\tau \to \nu_\tau +  \tau_\mathrm{jet}$ and these decays are performed using 
the program {\sc Tauola}~\cite{Golonka:2003xt} in order to
properly take into account the spin effects.
The resulting signature of the signal is thus:
$2 j + \tau_\mathrm{jet} + \mbox{$\not \!p_\perp$} \, .$
The dominant irreducible SM background arises from $W$ + 2 jets
production which we have simulated with help of the package 
ALPGEN \cite{Mangano:2002ea} again complemented with {\sc Tauola} 
to perform the $\tau$ decay. 

Our study is performed at parton level, without any parton showering or
hadronisation. Instead the momenta of the jets are smeared as a first 
approximation to take these, as well as detector effects, into account. 
After smearing the following basic cuts are applied:
$|\eta_{\tau_\mathrm{jet}}|<2.5$, $|\eta_j|<2.5$, $\Delta R_{jj}>0.4$, 
$\Delta R_{\tau_\mathrm{jet} j}>0.5$, and $p_{\perp \mathrm{jet}}>20$ GeV. 
We then apply the further cuts 
given in table~\ref{cuteffect} in order to suppress the background.
Here $m_\perp=
\sqrt{2p_{\perp\tau_\mathrm{jet}} \mbox{$\not \!p_\perp$}
[1-\cos(\Delta\phi)]}$,
with  $\Delta\phi$ being the azimuthal angle between $p_{\perp\tau_\mathrm{jet}}$ 
and \mbox{$\not \!p_\perp$}, is the transverse mass and 
$p_{\perp hj}$ ($p_{\perp sj}$) is the harder (softer) of the two jets. Although
we have not simulated the reducible QCD background explicitly, the cuts
\mbox{$\not \!p_\perp, p_{\perp\tau_\mathrm{jet}} >50 $ GeV} are primarily
included to take this into account. In order to get an estimate of the
sensitivity due to this choice we have also used an alternative set of harder
cuts, \mbox{$\not \!p_\perp, p_{\perp\tau_\mathrm{jet}} >100 $ GeV}.

Here we only report results in the  $m_h^\mathrm{max}$ scenario, for which we have used
$\mu=200$ GeV, $M_\mathrm{SUSY}=1$ TeV, $A_t=A_b=A_\tau=2$ TeV,
$M_2=200$ GeV, and $m_{\tilde g}=800$ GeV. The resulting signal cross-sections
for $\tan\beta=50$ and the two masses $m_{H^\pm}=175$ and 400 GeV are given in 
table~\ref{cuteffect} together with the irreducible SM background. The resulting number of
events and the significance $S/\sqrt{B}$ have been calculated using an integrated 
luminosity of 300 fb$^{-1}$ and a $\tau$ detection efficiency of 30\% .

\begin{table}[t]
\caption{The effect of the different cuts on the integrated cross-section for
  background  ($\sigma_{\rm b}$) and signal  ($\sigma_{\rm s}$) in the
  $m_h^\mathrm{max}$ scenario
  with $m_{H^\pm}=175$ and 400 GeV for $\tan\beta=50$.
}\centering
\begin{tabular}{c@{\qquad}c@{\qquad}ccc@{\qquad}ccc}
 \hline\hline
 & & \multicolumn{3}{c}{ $m_{H^\pm}=175$ GeV} &  
 \multicolumn{3}{c}{$m_{H^\pm}=400$ GeV}  \\
Cut [all in GeV] & $\sigma_{\rm b}$ (fb) & $\sigma_{\rm s}$ (fb) & 
$S$ & $S/\sqrt{B}$& $\sigma_{\rm s}$ (fb) & $S$ & $S/\sqrt{B}$ \\ \hline
Basic cuts &
 560000 & 55 & 4900  & 0.7 & 3.3 & 300 & 0.04 \\
$p_{\perp\tau_\mathrm{jet}}>50$ , $\mbox{$\!\not \!p_{\perp}$}>50$ &
 22000 & 25 & 2200 &  1.6 & 2.7 & 240 & 0.2 \\
$70 <m_{jj}<90$  &
 1700 & 21 & 1900 &  5 & 2.2 & 200 & 0.5 \\
$m_\perp>100$  &
 77 & 15 & 1400 &  16 & 2.1 & 190 & 2.3 \\
$p_{\perp hj}>50$  , $p_{\perp sj}>25$  &
 28 & 9.3 & 840 &  17 & 1.5 & 135 & 2.6 \\ 
 \hline\hline
\end{tabular}
\label{cuteffect} 
\end{table}

\begin{figure}[ht]
\centering
\includegraphics[width=6cm]{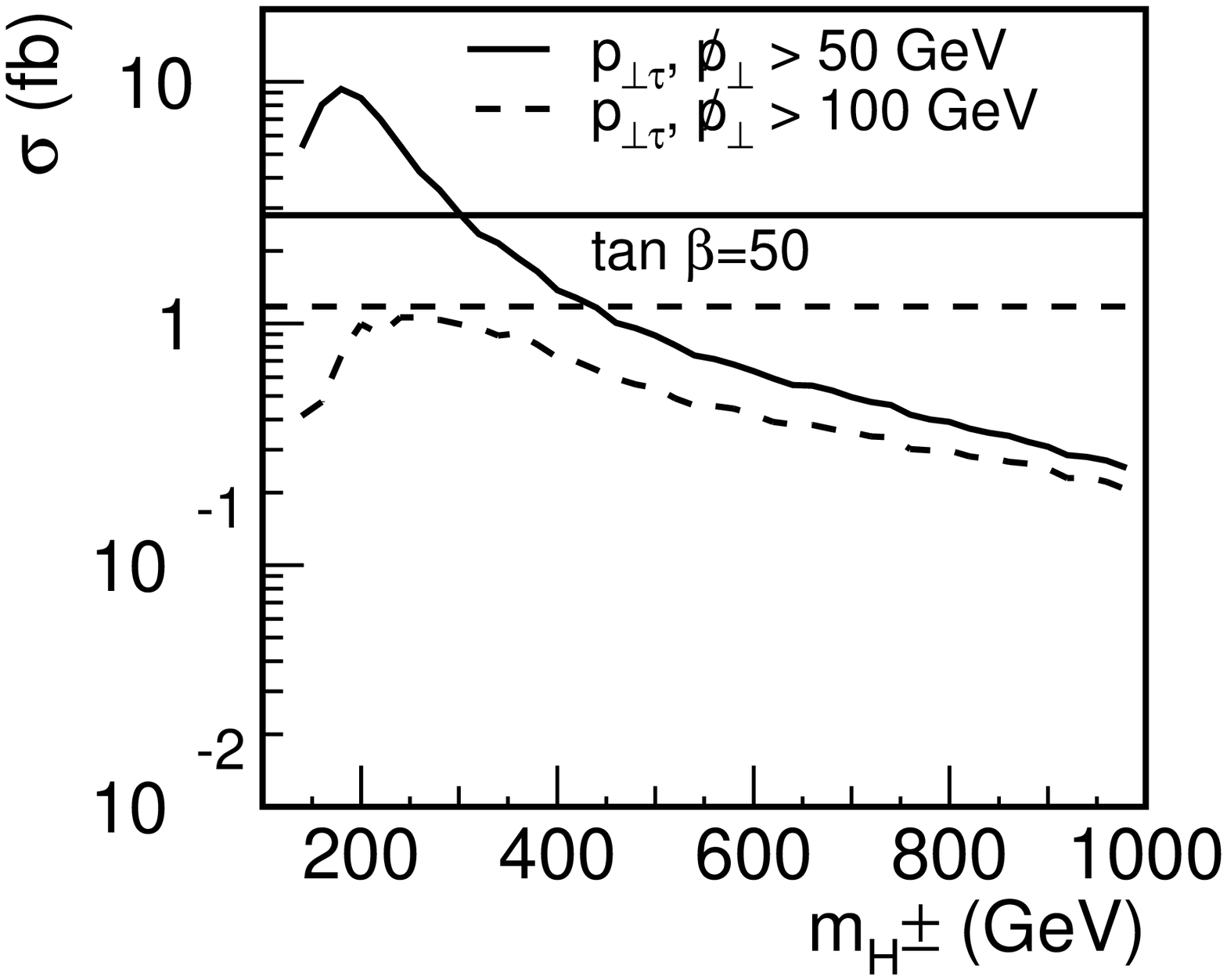} 
\hspace{1.5cm}
\includegraphics[width=6cm]{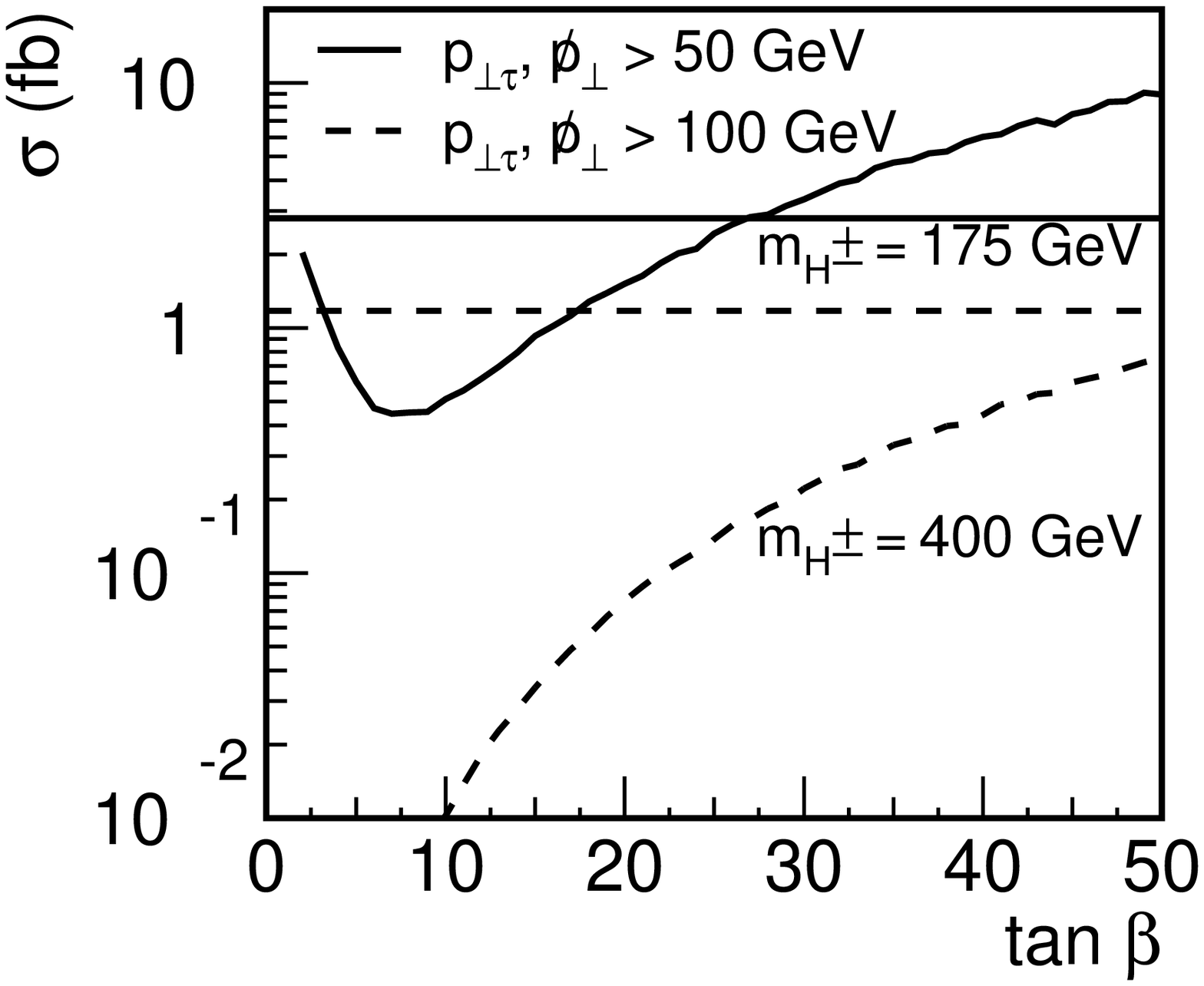}
\caption{$H^\pm$ mass and $\tan\beta$ dependence of the integrated
  cross-section in the $m_h^\mathrm{max}$ scenario. 
  Solid curves are with all cuts
  of table~\ref{cuteffect}, and dashed curves are with the harder  cuts 
  $p_{\perp\tau_\mathrm{jet}} , \!\not \!p_{\perp}>100$ GeV.
  The horizontal lines
  correspond to $\frac{S}{\sqrt{B}}=5$.
}\label{cutsigma_mhctanbeta} 
\end{figure}

The $m_{H^\pm}$ and $\tan\beta$ dependence of the cross-section after all cuts 
of table~\ref{cuteffect} are shown in figure~\ref{cutsigma_mhctanbeta}
as solid curves whereas dashed curves denote the cross-section
for the harder cuts 
$p_{\perp\tau_\mathrm{jet}}, \mbox{$\!\not \!p_{\perp}$}>100$ GeV.
The left plot is for $\tan \beta=50$ whereas
in the right plot we have
used $m_{H^\pm}=175$~GeV
(400~GeV) for the solid (dashed) line.
The horizontal lines indicate the cross-section needed for
$\frac{S}{\sqrt{B}}=5$, corresponding to
 $\tan\beta \gtrsim 30$ if $m_{H^\pm}=175$~GeV and 
$150~\textrm{GeV} \lesssim m_{H^\pm} \lesssim 300 $~GeV  if $\tan\beta =50$
with the softer cuts $p_{\perp\tau_\mathrm{jet}}, \mbox{$\!\not \!p_{\perp}$}>50$ GeV, 
whereas with the harder cuts
$\tan\beta$ has to be larger than at least 50.

\begin{figure}[ht]
\centering
\includegraphics[width=6cm]{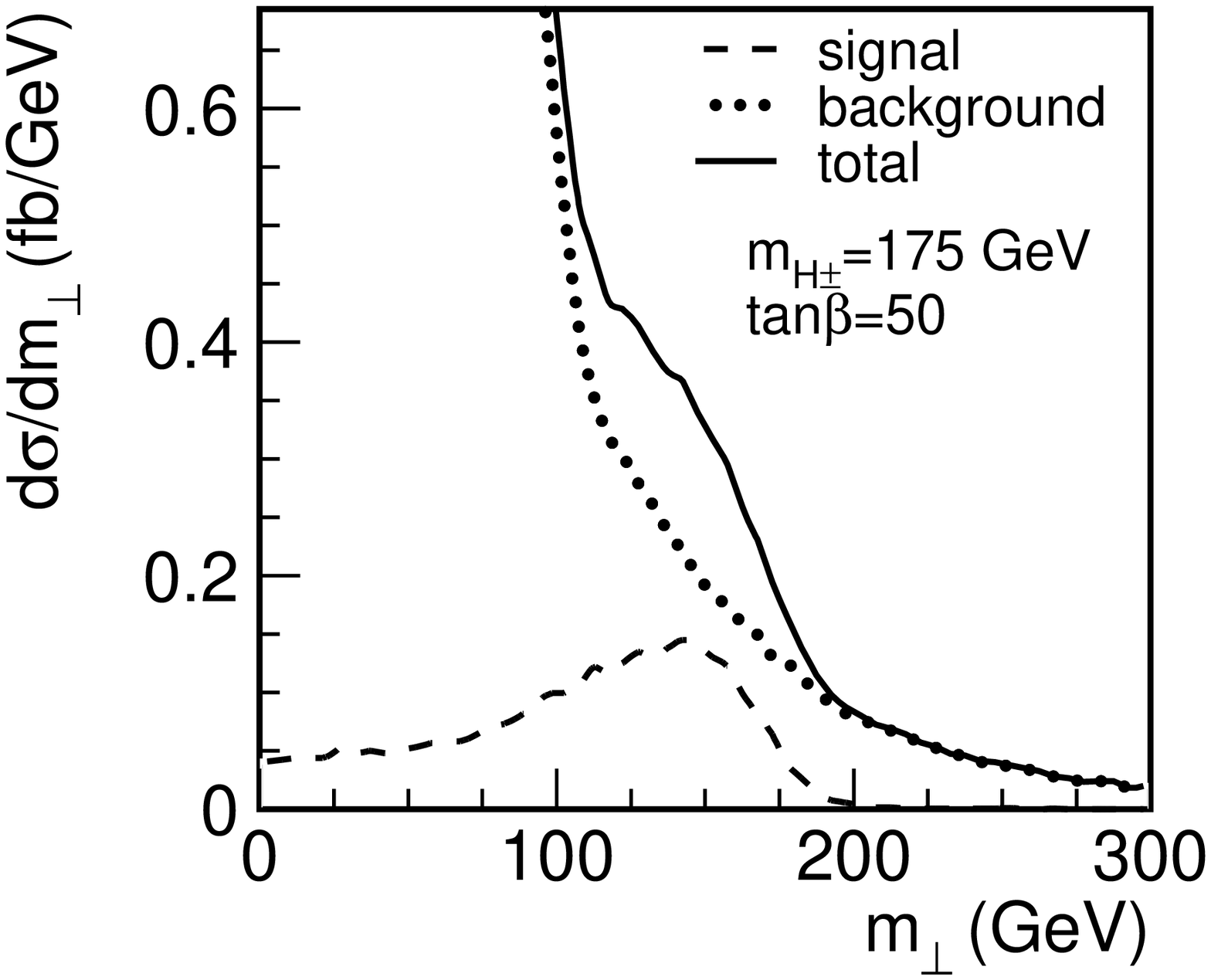}
\hspace{1.5cm}
\includegraphics[width=6cm]{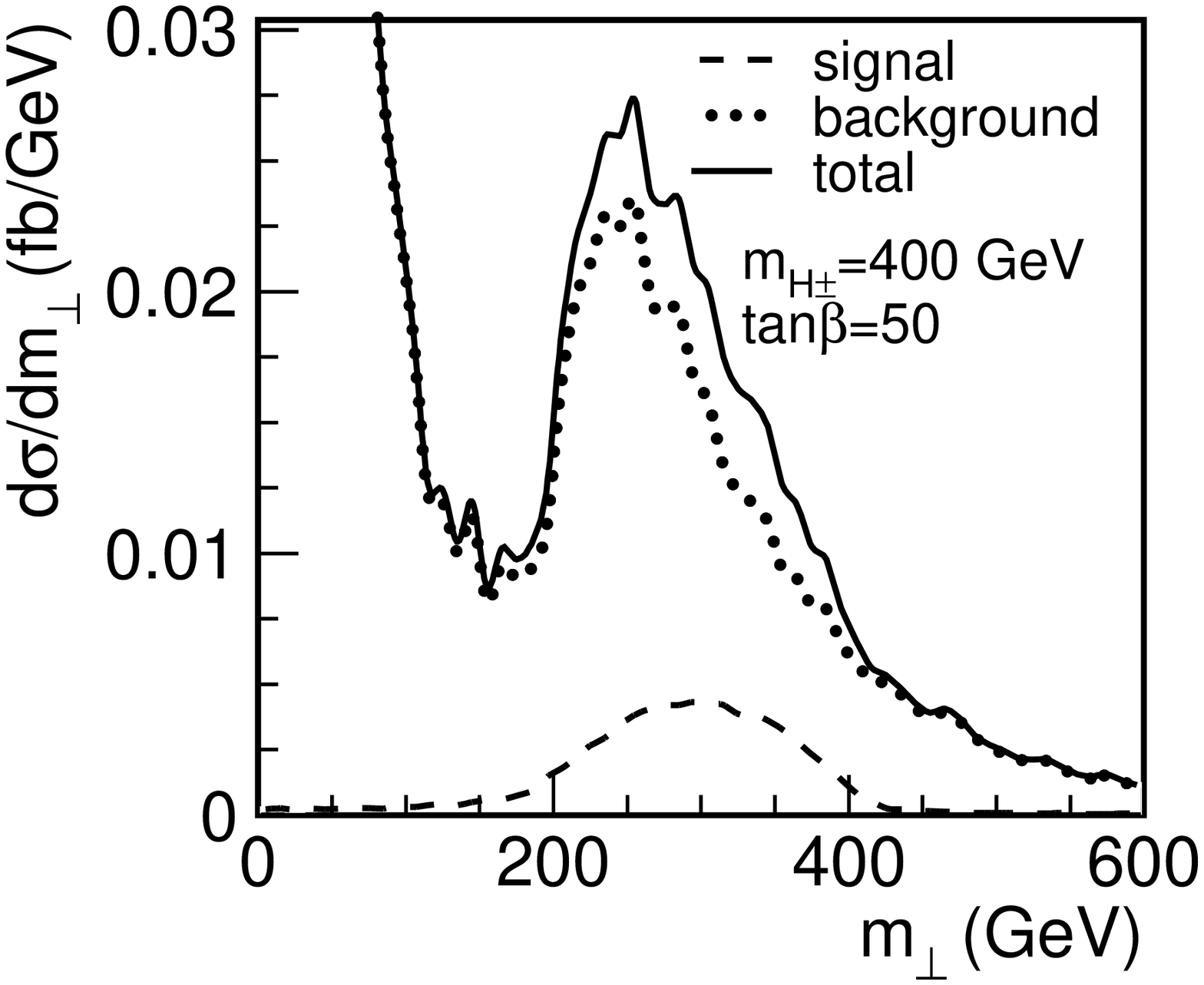}
\caption{The $m_\perp$ distribution for the signal (dashed) in the
  $m_h^\mathrm{max}$ scenario with $\tan\beta=50$ and $m_{H^\pm} = 175$ GeV
  (left) as well as $m_{H^\pm} = 400$ GeV (right) together with the 
  background (dotted) with all cuts of table~\ref{cuteffect}
  (for $m_{H^\pm} = 400$ GeV the cuts
  $p_{\perp\tau_\mathrm{jet}}, \mbox{$\!\not \!p_{\perp}$} > 100$~GeV are
  used). }\label{m175} 
\end{figure}

Figure~\ref{m175} shows the resulting 
$m_\perp$ distribution for $m_{H^\pm} = 175$~GeV as well as
$m_{H^\pm} = 400$~GeV in the case $\tan\beta =50$ compared to the background after 
all cuts in table~\ref{cuteffect} have been applied.
In the high mass case the harder cuts
$p_{\perp\tau_\mathrm{jet}}, \mbox{$\!\not \!p_{\perp}$} > 100$~GeV
are used giving $S/\sqrt{B}=3.2$.
Applying an upper cut  $m_\perp < 200$~GeV ($m_\perp < 500$~GeV)
for $m_{H^\pm} = 175$~GeV ($m_{H^\pm} = 400$~GeV)
only marginally improves $S/\sqrt{B}$ from 17 (3.2) to 19 (3.3).
In the same figure we also see that the harder cuts create a fake peak in the
background. Finally, using the harder cuts 
$p_{\perp\tau_\mathrm{jet}}, \mbox{$\!\not \!p_{\perp}$} > 100$~GeV the significance 
for $m_{H^\pm} = 175$ GeV and $\tan\beta =50$ is reduced to
 $S/\sqrt{B}=3.1$. However, in this case using an upper cut 
 $m_\perp < 200$~GeV is beneficial leading to a significance of 
 $S/\sqrt{B}=6.4$. For more details we refer to~\cite{Eriksson:2006yt}.

\ack
This work has been supported by the G\"oran Gustafsson Foundation.

\section*{References}

\bibliography{hep2007}

\end{document}